\documentclass[aps,prl,twocolumn,preprintnumbers,amsmath,amssymb,superscriptaddress]{revtex4} 
\usepackage{graphicx}
\usepackage{bm}
\usepackage[final]{pdfpages}

\begin{document}
\title{Anomalous Josephson effect in topological insulator-based Josephson trijunction}
\date{\today} 

\author{Xiang Zhang}
\affiliation{Beijing National Laboratory for Condensed Matter Physics, Institute of Physics, Chinese Academy of Sciences, Beijing 100190, China} \affiliation{School of Physical Sciences, University of Chinese Academy of Sciences, Beijing 100049, China}
\author{Zhaozheng Lyu}
\affiliation{Beijing National Laboratory for Condensed Matter Physics, Institute of Physics, Chinese Academy of Sciences, Beijing 100190, China}
\author{Guang Yang}
\affiliation{Beijing National Laboratory for Condensed Matter Physics, Institute of Physics, Chinese Academy of Sciences, Beijing 100190, China} \affiliation{School of Physical Sciences, University of Chinese Academy of Sciences, Beijing 100049, China}
\author{Bing Li}
\affiliation{Beijing National Laboratory for Condensed Matter Physics, Institute of Physics, Chinese Academy of Sciences, Beijing 100190, China} \affiliation{School of Physical Sciences, University of Chinese Academy of Sciences, Beijing 100049, China}
\author{Yan-Liang Hou}
\affiliation{Beijing National Laboratory for Condensed Matter Physics, Institute of Physics, Chinese Academy of Sciences, Beijing 100190, China} \affiliation{School of Physical Sciences, University of Chinese Academy of Sciences, Beijing 100049, China}
\author{Tian Le}
\affiliation{Beijing National Laboratory for Condensed Matter Physics, Institute of Physics, Chinese Academy of Sciences, Beijing 100190, China}
\author{Xiang Wang}
\affiliation{Beijing National Laboratory for Condensed Matter Physics, Institute of Physics, Chinese Academy of Sciences, Beijing 100190, China}\affiliation{School of Physical Sciences, University of Chinese Academy of Sciences, Beijing 100049, China}
\author{Anqi Wang}
\affiliation{Beijing National Laboratory for Condensed Matter Physics, Institute of Physics, Chinese Academy of Sciences, Beijing 100190, China}\affiliation{School of Physical Sciences, University of Chinese Academy of Sciences, Beijing 100049, China}
\author{Xiaopei Sun}
\affiliation{Beijing National Laboratory for Condensed Matter Physics, Institute of Physics, Chinese Academy of Sciences, Beijing 100190, China}\affiliation{School of Physical Sciences, University of Chinese Academy of Sciences, Beijing 100049, China}
\author{Enna Zhuo}
\affiliation{Beijing National Laboratory for Condensed Matter Physics, Institute of Physics, Chinese Academy of Sciences, Beijing 100190, China}\affiliation{School of Physical Sciences, University of Chinese Academy of Sciences, Beijing 100049, China}
\author{Guangtong Liu}
\affiliation{Beijing National Laboratory for Condensed Matter Physics, Institute of Physics, Chinese Academy of Sciences, Beijing 100190, China}\affiliation{Huairou Division, Institute of Physics, Chinese Academy of Sciences, Beijing 100190, China}\affiliation{Songshan Lake Materials Laboratory, Dongguan, Guangdong 523808, China}
\author{Jie Shen}
\affiliation{Beijing National Laboratory for Condensed Matter Physics, Institute of Physics, Chinese Academy of Sciences, Beijing 100190, China}\affiliation{Huairou Division, Institute of Physics, Chinese Academy of Sciences, Beijing 100190, China}
\author{Fanming Qu}
\affiliation{Beijing National Laboratory for Condensed Matter Physics, Institute of Physics, Chinese Academy of Sciences, Beijing 100190, China} \affiliation{Huairou Division, Institute of Physics, Chinese Academy of Sciences, Beijing 100190, China}\affiliation{Songshan Lake Materials Laboratory, Dongguan, Guangdong 523808, China}
\author{Li Lu} \email[Corresponding authors: ]{lilu@iphy.ac.cn}
\affiliation{Beijing National Laboratory for Condensed Matter Physics, Institute of Physics, Chinese Academy of Sciences, Beijing 100190, China} \affiliation{School of Physical Sciences, University of Chinese Academy of Sciences, Beijing 100049, China} \affiliation{Huairou Division, Institute of Physics, Chinese Academy of Sciences, Beijing 100190, China}\affiliation{Songshan Lake Materials Laboratory, Dongguan, Guangdong 523808, China}


\begin{abstract}
We studied anomalous Josephson effect (AJE) in Josephson trijunctions fabricated on Bi$_2$Se$_3$, and found that the AJE in T-shaped trijunctions significantly alters the Majorana phase diagram of the trijunctions, when an in-plane magnetic field is applied parallel to two of the three single junctions. Such a phenomenon in topological insulator-based Josephson trijunction provides unambiguous evidence for the existence of AJE in the system, and might provide an additional knob for controlling the Majorana bound states in the Fu-Kane scheme of topological quantum computation. 
\end{abstract}

\maketitle

With breaking both time-reversal symmetry and chiral symmetry, the Andreev reflection processes in a superconductor/normal-state metal/superconductor (S/N/S) type of Josephson junction will accumulate an additional phase \cite{1}, such that the ground state of the Josephson junction will no longer be located at phase difference $\varphi=0$ or $\varphi=\pi$ as in a usual Josephson junction, but shifted to a finite value $\varphi=\varphi_0$. This phenomenon is known as the AJE \cite{1,2,3}, and has been studied theoretically in Josephson junctions containing either magnetic elements [2], or based on topological insulators \cite{4,5}, silicene \cite{6} and other materials with strong spin-orbit coupling \cite{3,7,8,9}. The anomalous phase shift could cause spin-galvanic effect, also known as inverse Edelstein effect \cite{10,11}. Such a magneto-electric effect might be useful in future superconducting electronics and spintronics \cite{12}, such as to drive quantum phase batteries \cite{13}, to form superconducting quantum memories \cite{14,15}, or to induce topological superconductivity \cite{16,17,18,19}, etc.

From the experimental side, however, very limited investigations on AJE have been reported, so far only on two-terminal devices (i.e., single Josephson junctions) made of InAs-based quantum dots \cite{20}, InAs nanowires \cite{21}, InAs two-dimensional electron gas \cite{22} and three-dimensional topological insulator (3D TI) Bi$_2$Se$_3$ \cite{23}. In these two-terminal devices, the application of an in-plane magnetic field not only causes an anomalous phase shift but often a normal phase shift due to the existence of an out-of-plane component of the field, because of slight but practically unavoidable misalignment in the field direction. The normal phase shift caused by the out-of-plane component mixes up with the anomalous phase shift, disturbing the AJE phenomenon. In order to identify the existence of AJE, control experiments are often needed such as to use electric gating to merely adjust the anomalous phase shift via spin-orbit coupling \cite{20,23}.

\begin{figure*}
\includegraphics[width=1 \linewidth]{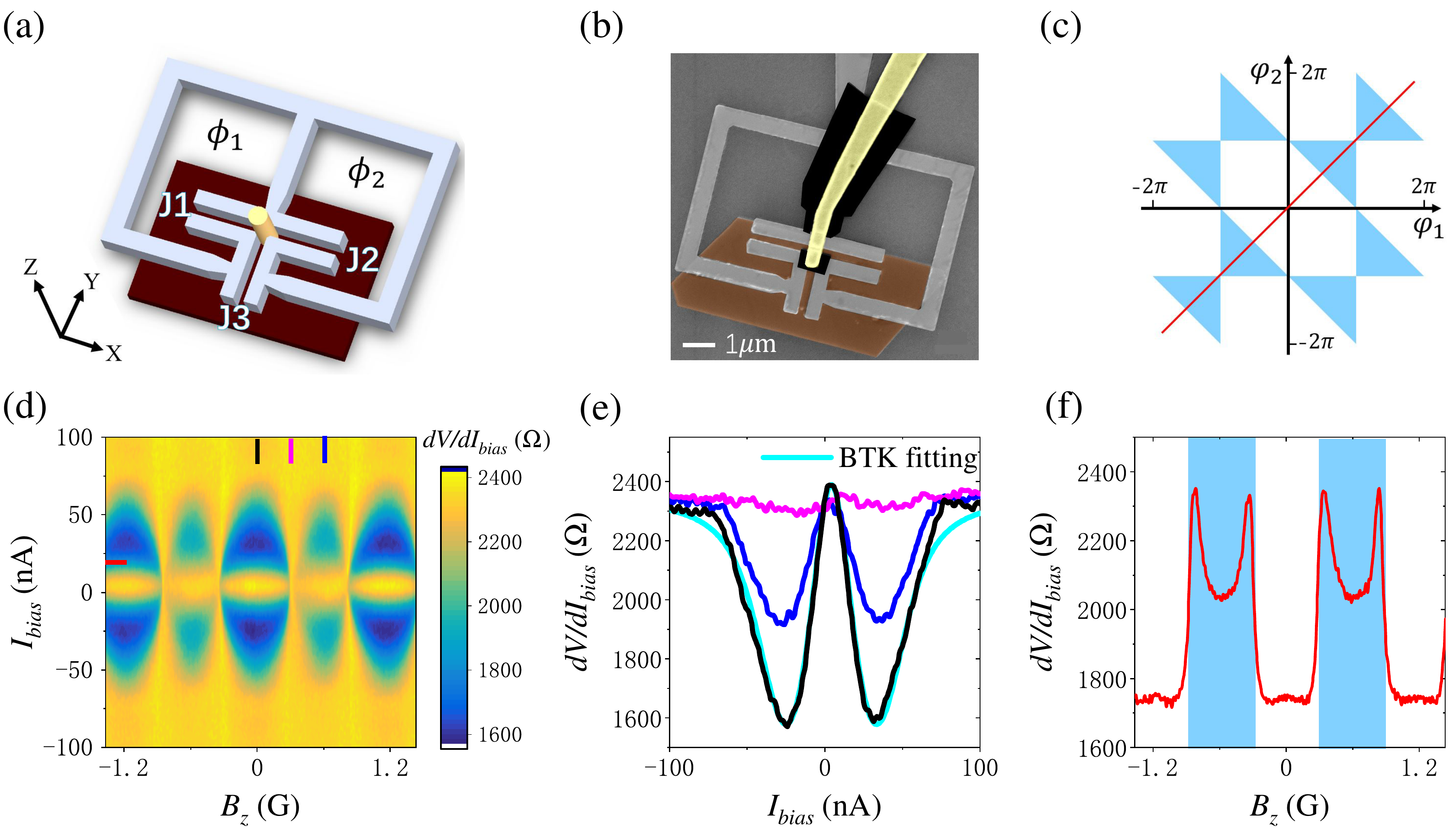}
\caption{\label{fig:fig1} {The Josephson trijunction and the method of measuring the minigap at its center. (\textbf{a}) Schematic of the device on the upper surface of a 3D TI (in puce), with two superconducting loops (each with an area of $S$) connecting to single junctions J1 and J2 along the x-direction are for the purpose to adjust the phase differences across these single junctions via magnetic flux $\phi_1$ and $\phi_2$, respectively. A normal-state metal electrode (in yellow) is used for detecting the minigap at the center of the trijunction via contact resistance measurement. (\textbf{b}) False-colored scanning electron microscopy image of the trijunction device constructed on the surface of a nanoplate of Bi$_2$Se$_3$ single crystal. (\textbf{c}) The Majorana phase diagram of the trijunction in the parametric space spanned by $\varphi_1$ and $\varphi_2$. The blue regions represent the minigap-closing state, and the white regions represent the minigap-opening state \cite{24,25}. The red line indicates the trace of the trijunction's state during sweeping global magnetic field along the out-of-plane direction. (\textbf{d}) The contact resistance $dV/dI_{bias}$ as functions of the bias current $I_{bias}$ and the magnetic field $B_z$ along the out-of-plane direction. (\textbf{e}) The vertical line cuts in (d) at magnetic fields indicated by the marks with corresponding colors in (d). The cyan line is the BTK fitting to the data of the fully minigap-opening state (black line). (\textbf{f}) The horizontal line cut in (d) at bias current of $\sim$20 nA. The lowest value of the $dV/dI_{bias}$ represents the fully minigap opening state, the peak value represents the minigap closing state, and the intermediate values correspond to the minigap reopening state \cite{25}.}}
\end{figure*}

In this experiment, we study the AJE in Josephson trijunction based on topological insulators, and demonstrate that the anomalous phase shift alters the response of the trijunction in out-of-plane magnetic field, by which the happening of AJE can be unambiguously identified. Such an anomalous phase shift in Josephson trijunctions, which can be manipulated by turning the strength and direction of the in-plane magnetic field, may provide an additional knob for controlling the Majorana bound states (MBSs) in the Fu-Kane scheme of topological quantum computation \cite{24}.

The trijunctions used in this experiment were first proposed by L. Fu and C. L. Kane in 2008 for the purpose of hosting MBSs \cite{24}. In 2015, S. Vijay and L. Fu further proposed that by employing a surface code technique, universal quantum computing can be implemented by using the trijunctions. Later on, G. Yang, Z. Z. Lyu and coworkers successfully fabricated Y-shaped Josephson trijunction on the surface of topological insulators \cite{25}. By introducing two superconducting loops, the authors were able to adjust the phase differences in each single junctions individually via applying local magnetic flux, by which they verified the 2D Majorana phase diagram of Josephson trijunctions predicted by Fu and Kane. In the present experiment, we fabricated T-shaped Josephson trijunction on the surface of topological insulator. The reason for choosing the T shape rather than the Y shape is for better studying the in-plane angle dependence of the magnetic field on AJE. We expect no difference in their responses to the out-of-plane magnetic field between the two shapes of devices.

Illustrated in Fig. 1(a) is a T-shaped trijunction constructed on the surface of a 3D TI. And shown in Fig. 1(b) is the scanning electron microscopy image of the real device used in this experiment, which is constructed on a flake of Bi$_2$Se$_3$ single crystal ($\sim$30 nm in thickness). The device contains three single junctions J1, J2 and J3, with J1 and J2 aligned along the x-direction and J3 aligned along the y-direction. Each Al/Bi$_2$Se$_3$/Al single junction is a proximitized Josephson junction with Andreev bound states (ABSs) in the Bi$_2$Se$_3$ part between the Al electrodes. Depending on the phase differences $\varphi_1$ and $\varphi_2$ in J1 and J2, which can be controlled by the magnetic flux $\phi_1$ and $\phi_2$ in the two superconducting loops, there may exist localized MBSs at the center of the trijunction. The appearance of MBSs is manifested as the closure of local minigap (defined as the energy spacing between the lowest-energy electron-like and hole-like Andreev levels). According to previous theoretical and experimental studies, the minigap at the center of the trijunction varies periodically in the phase space spanned by $\varphi_1$ and $\varphi_2$. It is fully open in the white areas of the Fu-Kane’s Majorana phase diagram (Fig. 1(c)), and undergoes completely closing and slightly reopening in the blue areas of the phase diagram \cite{25}. The status of the minigap can be detected through measuring the contact resistance of a normal-state metal electrode (Au) connected to the Bi$_2$Se$_3$ surface at the center of the trijunction, as illustrated in yellow in Figs. 1(a) and 1(b). Other than at the center, the Au electrode is electrically isolated with the rest part of the device by the surface oxidation layer of the Al film and also by a layer of overexposed polymethyl methacrylate (in black in Fig. 1(b)).

\begin{figure*}
\includegraphics[width=1 \linewidth]{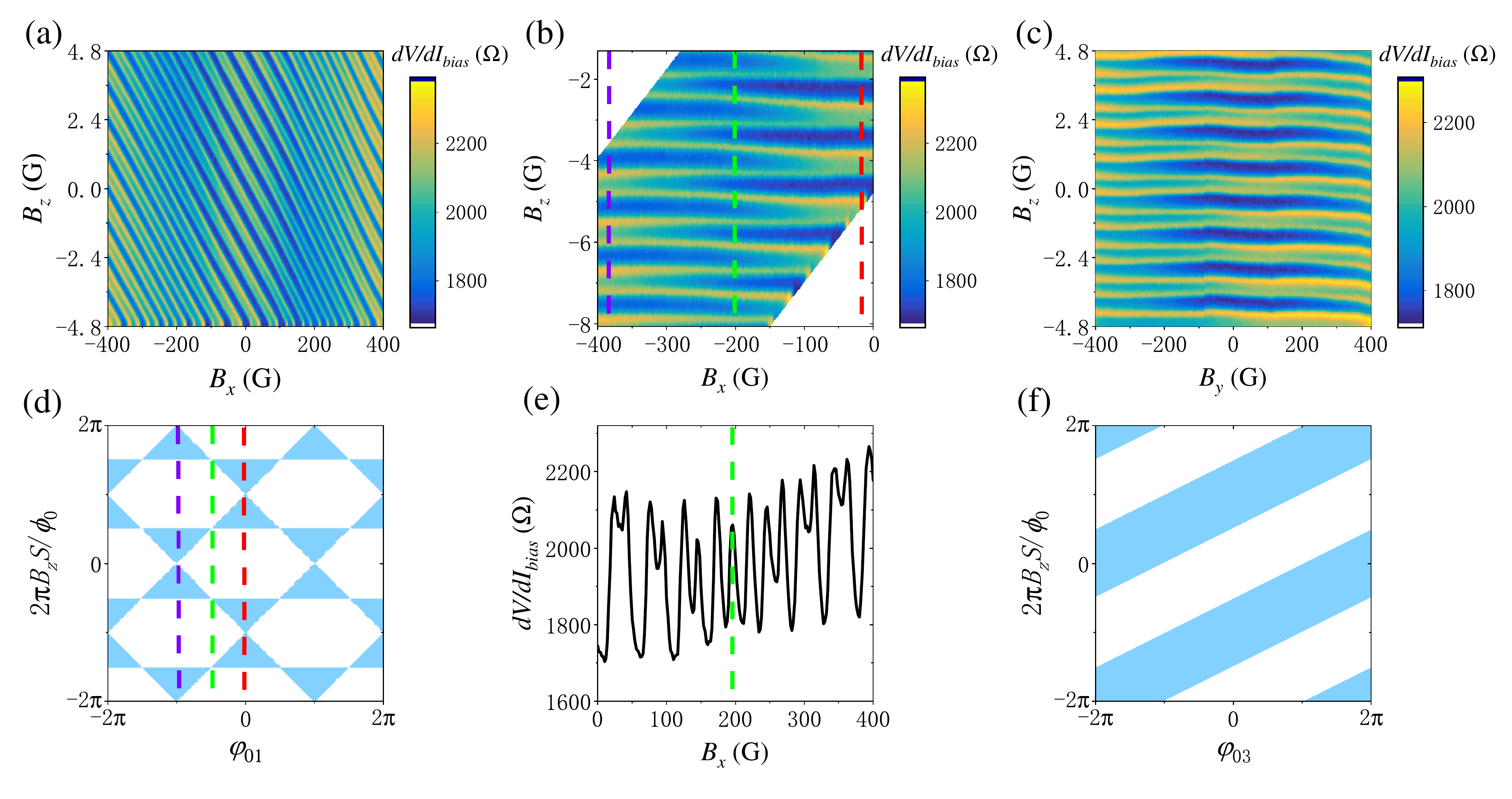}
\caption{\label{fig:fig2} {The contact resistance $dV/dI_{bias}$ of the probe electrode as functions of the out-of-plane magnetic field $B_z$ and the in-plane magnetic field along the x and y directions. (\textbf{a}) The $dV/dI_{bias}$ as functions of $B_x$ and $B_z$. The tilting of the patterns is caused by the unintentionally existed out-of-plane component of the in-plane magnetic field. (\textbf{b}) A part of the pattern in (a) after subtracted the tilted background. (\textbf{c}) The $dV/dI_{bias}$ as functions of $B_y$ and $B_z$. (\textbf{d}) The Majorana phase diagram in Fig. 1(c) after rotated by 45$^{\circ}$ counterclockwise, forming a new Majorana phase diagram in the parametric space spanned by $\varphi_{01}$ and 2$\pi B_z S/\phi_0$. The change of the trijunction's state along the dashed lines explains the data along the dashed lines in (b) with corresponding colors. (\textbf{e}) The $dV/dI_{bias}$ as a function of $B_x$ along the horizontal line cut in (a) at $B_z$=0. The anomalous phase shift reaches $\sim \pi/2$ at the green dashed line. (\textbf{f}) The expected Majorana phase diagram in parametric space spanned by 2$\pi B_z S/\phi_0$ and $\varphi_{03}$, explaining the measured data in (c).}}
\end{figure*}

For obtaining the contact resistance of the Au electrode, we performed three-terminal electron transport measurements as described in Refs. 25 and 26, in which only the voltage drop across the electrode that shared by the voltage measurement loop and the current loop was detected, by using low-frequency lock-in amplifier technique. In order to lower the electron temperature of the device and to increase the energy resolution, all electric leads were equipped with $\pi$ filters at room temperature, RC filters at $\sim$1 K, additional RC and copper powder filters at the base temperature in the dilution refrigerator.

Shown in Fig. 1(d) is the 2D map of the measured contact resistance as functions of the bias current $I_{bias}$ and the out-of-plane magnetic field $B_z$. And shown in Figs. 1(e) and 1(f) are the vertical and horizontal line cuts of the 2D map, respectively. The data in black, purple and blue colors in Fig. 1(e) represents the fully opening, completely closing, and slightly reopening states of the minigap at the center of trijunction, respectively. In Fig. 1(f), the lowest value of the $dV/dI_{bias}$ represents the fully opening state, the peak value represents the completely closing state, and the intermediate values correspond to the reopening state of the minigap.

The data of contact resistance can be quantitatively analyzed by using the Blonder-Tinkham-Klapwijk (BTK) formalism. The BTK theory was originally developed to describe the electron transport processes across the interface between a normal-state metal and a superconductor \cite{27}. Previously, we have shown that the BTK formalism can be successfully applied to analyze the normal-state metal and TI interface in the junction area, by which the information of local minigap can be obtained \cite{25,27}. In Fig. 1(e), the cyan line is the BTK fitting to the black line which were measured in zero magnetic field, corresponding to the fully minigap-opening state. The fitting parameters are: the minigap $\Delta_{mini}\sim$ 57$\mu$eV, and the interfacial barrier strength $Z\sim 0.63$. For the detailed method of the fitting please refer to Refs. 25 and 27.

After understanding the basic features of the device, we then studied the contact resistance of the probe electrode as functions of $B_x$ and $B_z$. The results are shown in Fig. 2(a). Due to the unintentionally existed out-of-plane component of the in-plane magnetic field, the pattern is often tilted. Based on the slope of the pattern, we obtain that the misalignment angle of the substrate is about 1.25$^{\circ}$. Shown in Fig. 2(b) is a part of the pattern in Fig. 2(a) after the tilted background is removed. The yellow lines in the 2D plots represent the minigap closing state, whereas the dark blue and the light blue regions between the yellow lines represent the fully minigap opening and minigap reopening states, respectively. From Figs. 2(a) and 2(b) it appears that with varying $B_x$, two neighboring yellow lines undergo a process of splitting apart and recombining with their other neighbors. This is a unique signature observed on T-shaped Josephson trijunction, providing unambiguous evidence for the existence of an anomalous phase shift.

Since the mean free path of electron in our junctions, as reported earlier \cite{28}, is smaller than the length of the junctions, our junctions are in the diffusive regime. The anomalous phase shift $\varphi_{0i}$ takes the form of $\varphi_{0i}=\tau m^{*2} E_Z (\alpha L)^3/(3\hbar^6 D)$ when the magnetic field is applied parallel to the $i^{th}$ single junction \cite{7}, where $\tau$ is the elastic scattering time, $m^*$ is the effective electron mass, $E_Z$ is the Zeeman energy, $\alpha$ is the is the Rashba spin-orbit coupling coefficient, $L$ is the length of junction and $D$ is the diffusion constant. When the in-plane magnetic field is applied along the x direction, it generates anomalous phase shifts in junctions J1 and J2, with $\varphi_{01}=-\varphi_{02}$ along a given clockwise or counterclockwise loop direction. In J3 the effect is negligible because the in-plane field is perpendicular to the junction. The total phase shifts in the three single junctions are:
\begin{align*}
&\varphi_1=2\pi B_zS/\phi_0+\varphi_{01}\\
&\varphi_2=2\pi B_zS/\phi_0-\varphi_{01}\\
&\varphi_3=-\varphi_1-\varphi_2=-4\pi B_zS/\phi_0 
\end{align*}

It can be seen that there are two independent phase differences $\varphi_1$ and $\varphi_2$ controlled by $\varphi_{01}$ and the magnetic flux $B_z S$ in the superconducting loops. The parametric space spanned by $\varphi_{01}$ and $B_z S$, as shown in Fig. 2(d), is actually the same as the one for displaying the Majorana phase diagram (i.e., Fig. 1(c)), except for being rotated by 45$^{\circ}$ counterclockwise. This is because that $(\varphi_1-\varphi_2)/2=\varphi_{01}$ and $(\varphi_1+\varphi_2)/2=-2\pi B_zS/\phi_0$. From Fig. 2(d) it can clearly be seen that applying $B_z$ causes merely vertical shift of the trijunctions's state in this new Majorana phase diagram, and applying in-plane magnetic field causes merely horizontal shift in the new Majorana phase diagram, purely because of the anomalous phase shift. With sweeping both $B_z$ and $B_x$, and with the later has an out-of-plane component, it yields typically the data shown in Fig. 2(a). The tilting of the pattern is simply caused by the out-of-plane component of $B_x$. 

The existence of an anomalous phase shift can also be clearly seen from the line cut of the 2D plot at $B_z$=0, as shown in Fig. 2(e). It shows that, with sweeping $B_x$, the fully minigap opening state is gradually converted to the minigap reopening state. At the green dashed line the conversion is half way down, corresponding to an anomalous phase shift of $\sim \pi$/2 

In the next, we measured the contact resistance of the probe electrode as functions of $B_y$ and $B_z$. The results are shown in Fig. 2(c). It happened that the contribution of the out-of-plane component of the in-plane magnetic field to the tilting of the pattern was cancelled by the anomalous phase in J3, so that the pattern in Fig. 2(c) is almost leveled. It can be seen that the anomalous phase shift does not effectively cause the yellow lines of the pattern to alternatively split and combine with other neighbors as is shown in Figs. 2(a) and 2(b). 

When the in-plane magnetic field is applied along the y direction, it generates an anomalous phase shift $\varphi_{03}$ in junction J3, leaving J1 and J2 untouched. The total phase shifts in the three single junctions are:
\begin{align*}
&\varphi_1=2\pi B_zS/\phi_0\\
&\varphi_2=2\pi B_zS/\phi_0\\
&\varphi_3=-4\pi B_zS/\phi_0+\varphi_{03} 
\end{align*}

In this case, J3 together with the superconducting loops form a $\varphi_{03}$-loop rf SQUID, multiply connected with another $\varphi_{03}$-loop formed by the other two single junctions. Spontaneous supercurrent will be generated in these loops by the anomalous phase shift in J3. Also, spontaneous magnetic flux will be created in these loops, and the magnetic flux will be significant if both the loop inductance and the critical Josephson supercurrents of the junctions are big enough. These spontaneous generated supercurrent and magnetic flux will cause normal phase shifts on J1 and J2. However, their effects will be exactly canceled out along the $\varphi_1-\varphi_2$ direction in phase space. Therefore, sweeping $B_y$ does not cause horizontal shift of the trijunction’s state in the new Majorana phase diagram (Fig. 2(d)), but merely causing vertical shift there, leading to the pattern illustrated in Fig. 2(f). This accounts for the measured data shown in Fig. 2(c). 

\begin{figure}
\includegraphics[width=1 \linewidth]{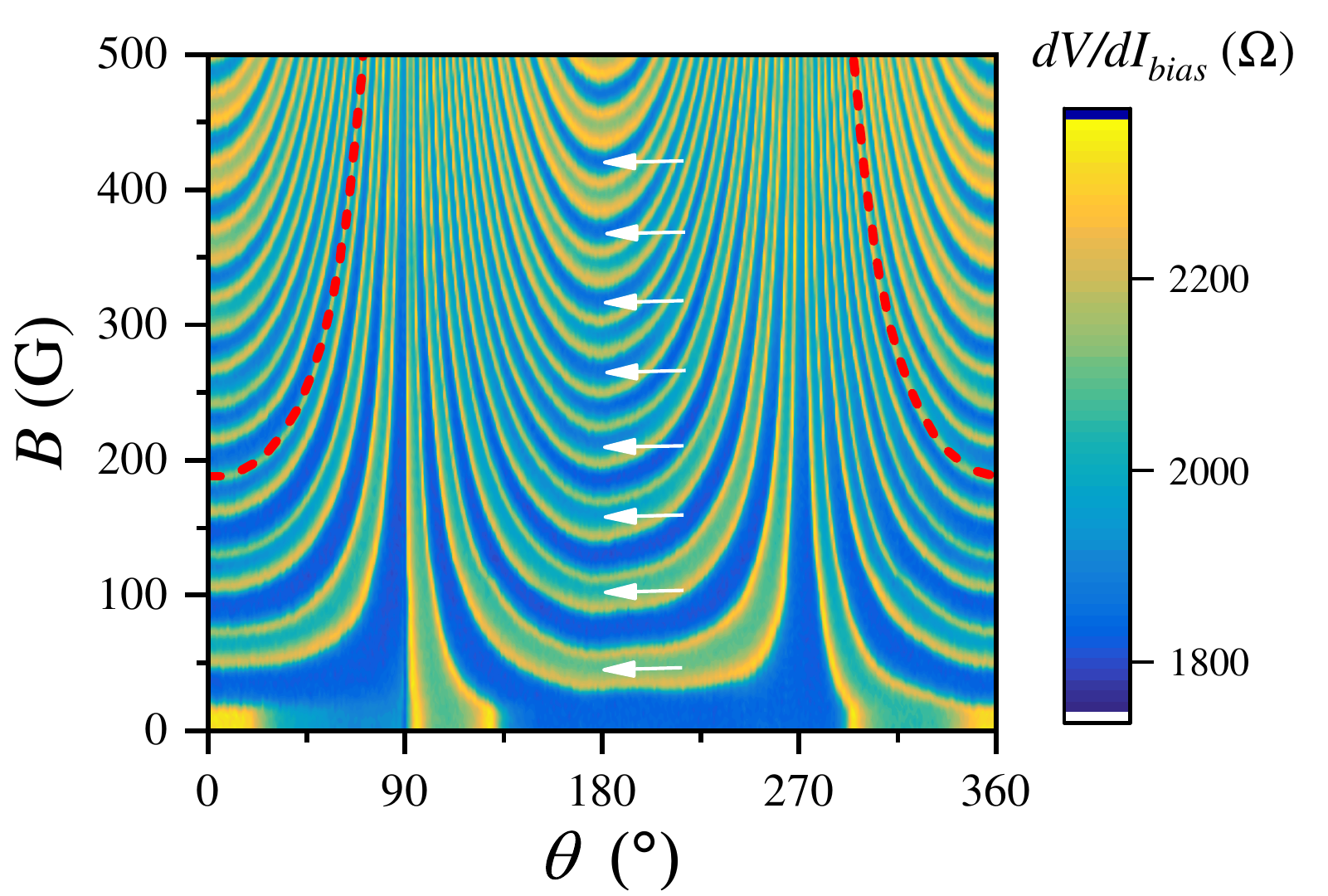}
\caption{\label{fig:fig3} {The contact resistance $dV/dI_{bias}$ as functions of the strength and angle of the in-plane magnetic field in the absence of $B_z$. The ``in-plane'' magnetic field has a small out-of-plane component, leading to the appearance of multiple strips curving in a functional form of $1/\cos\theta$ (illustrated by the red dotted lines), where $\theta$ is the angle of the in-plane magnetic field with respect to the x direction. The evenly spaced arrows are used to indicate and to trace from the bottom of the green-colored stripes (corresponding to the minigap reopening states) at $\theta$=180$^{\circ}$. It shows that the arrowed positions gradually turn blue (corresponding to the minigap opening states), implying that a $\pi/2$ phase shift is gradually implemented from the bottom to the top of the frame, due to the accumulation of the anomalous phase shift.}}
\end{figure}

We have also studied the anomalous phase shift with rotating the direction of the in-plane magnetic field. Shown in Fig. 3 is the contact resistance as functions of the magnitude and angle of the in-plane magnetic field. Let us first look at the vertical line cut at $\sim$180$^{\circ}$. With increasing the magnitude of the in-plane magnetic field, the yellow lines of the pattern, which represent the completely closing state of the minigap, undergo splitting and recombining processes with the accumulation of the anomalous phase shift. Usually, the rotating plane of the magnetic field is not fully parallel to the substrate plane of the device, but with a misalignment angle which maximized at certain in-plane directions. For the present device, this angle happens to maximize near $\theta$=0$^{\circ}$ or 180$^{\circ}$. The out-of-plane component of the magnetic field leads to the appearance of multiple strips near $\theta$=0$^{\circ}$ or 180$^{\circ}$, as shown in Fig. 3. The shape of the strips simply follows the functional form of $1/\cos\theta$ illustrated by the red dotted lines in Fig. 3. It can be seen that, even if there are misalignments in the directions of the in-plane and out-of-plane magnetic fields, the existence of AJE in Josephson trijunction can unambiguously be identified from the overall pattern of the contact resistance in this rotating field experiment.

To summarize, we have shown that the anomalous phase shift in “T”-shaped Josephson trijunction exhibits strong anisotropy with the direction of in-plane magnetic field, leading to unique modifications on the Majorana phase diagram which was originally proposed for the trijunction’s states in out-of-plane magnetic field. Such modifications can be unambiguously identified even if the in-plane magnetic field has unintentionally an out-of-plane component. The existence of an out-of-plane component actually helps the identification of the AJE. The anomalous phase shift might be used as a knob for controlling the MBSs in Josephson trijunctions \cite{24} and other proposed multi-terminal topological quantum devices \cite{29,30}, for the purposes of studying non-Abelian statistics and fault-tolerant topological quantum computing. This effect can also be used to study the in-plane anisotropy of the g factor of the materials that the proximitized Josephson junctions are constructed on \cite{31,32}.

This work was supported by the National Basic Research Program of China through MOST Grants No. 2016YFA0300601, No. 2017YFA0304700, and No. 2015CB921402; by the NSF China through Grants No. 11527806, No. 92065203, No. 12074417, No. 11874406, and No. 11774405; by the Beijing Academy of Quantum Information Sciences, Grant No. Y18G08; and by the Strategic Priority Research Program B of Chinese Academy of Sciences, Grants No.XDB33010300, No. XDB28000000 and No. XDB07010100; by the Synergetic Extreme Condition User Facility sponsored by the National Development and Reform Commission.

\end{document}